\documentclass[sigconf]{acmart}
\usepackage{graphicx}
\usepackage{caption}
\usepackage{subcaption}
\usepackage{hyphenat}
\usepackage{tikz}
\def\checkmark{\tikz\fill[scale=0.4](0,.35) -- (.25,0) -- (1,.7) -- (.25,.15) -- cycle;}
\usepackage{multicol}
\usepackage{setspace}
\usepackage{lipsum}

\AtBeginDocument{%
  \providecommand\BibTeX{{%
    \normalfont B\kern-0.5em{\scshape i\kern-0.25em b}\kern-0.8em\TeX}}}

\copyrightyear{2022}
\acmYear{2022}
\setcopyright{acmlicensed}\acmConference[L@S '22]{Proceedings of the Ninth ACM Conference on Learning @ Scale}{June 1--3, 2022}{New York City, NY, USA}
\acmBooktitle{Proceedings of the Ninth ACM Conference on Learning @ Scale (L@S '22), June 1--3, 2022, New York City, NY, USA}
\acmPrice{15.00}
\acmDOI{10.1145/3491140.3528274}
\acmISBN{978-1-4503-9158-0/22/06}

\begin{document}

\title[When Gamification Spoils Your Learning]{When Gamification Spoils Your Learning: A Qualitative Case Study of Gamification Misuse in a Language-Learning App}

\author{Reza Hadi Mogavi}
\authornote{Corresponding authors}
\affiliation{%
  \institution{HKUST}
  \country{Hong Kong SAR}
}
\email{rhadimogavi@cse.ust.hk}

\author{Bingcan Guo}
\authornote{Both co-authors contributed equally to this research.}
\affiliation{%
  \institution{HKUST}
  \country{Hong Kong SAR}
}
\email{bguoac@connect.ust.hk}

\author{Yuanhao Zhang}
\authornotemark[2]
\affiliation{%
  \institution{HKUST}
  \country{Hong Kong SAR}
}
\email{yzhangiy@connect.ust.hk}

\author{Ehsan-Ul Haq}
\affiliation{%
  \institution{HKUST}
  \country{Hong Kong SAR}
}
\email{euhaq@connect.ust.hk}

\author{Pan Hui}
\authornotemark[1]
\affiliation{%
  \institution{HKUST \& University of
Helsinki}
  \country{Hong Kong SAR \& Finland}
}
\email{panhui@cse.ust.hk}

\author{Xiaojuan Ma}
\authornotemark[1]
\affiliation{%
  \institution{HKUST}
  \country{Hong Kong SAR}
}
\email{mxj@cse.ust.hk}
%
\renewcommand{\shortauthors}{Reza Hadi Mogavi, et al.}

\begin{abstract}
More and more learning apps like Duolingo are using some form of gamification (e.g., badges, points, and leaderboards) to enhance user learning. However, they are not always successful. \textit{Gamification misuse} is a phenomenon that occurs when users become too fixated on gamification and get distracted from learning. This undesirable phenomenon wastes users' precious time and negatively impacts their learning performance. However, there has been little research in the literature to understand gamification misuse and inform future gamification designs. Therefore, this paper aims to fill this knowledge gap by conducting the first extensive qualitative research on gamification misuse in a popular learning app called Duolingo. Duolingo is currently the world's most downloaded learning app used to learn languages. This study consists of two phases: (I) a \textit{content analysis} of data from Duolingo forums (from the past nine years) and (II) \textit{semi-structured interviews} with 15 international Duolingo users. Our research contributes to the Human-Computer Interaction (HCI) and Learning at Scale (L@S) research communities in three ways: (1) elaborating the ramifications of gamification misuse on user \textit{learning}, \textit{well-being}, and \textit{ethics}, (2) identifying the most common reasons for gamification misuse (e.g., \textit{competitiveness}, \textit{overindulgence in playfulness}, and \textit{herding}), and (3) providing designers with practical suggestions to prevent (or mitigate) the occurrence of gamification misuse in their future designs of gamified learning apps.
\end{abstract}

\ccsdesc[500]{Human-centered computing~Empirical studies in HCI}
\ccsdesc[500]{Applied computing~Education}
\ccsdesc[500]{Applied computing~Computer games}
\keywords{Student-centered education, Duolingo, gamified education, gamification, learning app, misuse, HCI, L@S, qualitative research.}


\maketitle

\section{Introduction}
In recent years, the use of gamification for educational purposes has raised in popularity \cite{goethe2019gamification, 10.1145/3338286.3344394, Stockinger2014, 10.1145/3290605.3300361}. Simply put, gamification is a motivational technique that uses game design elements (such as badges, points, and leaderboards) in non-game contexts to trigger positive user behaviors \cite{10.1145/2181037.2181040, goethe2019gamification}. The main objective of using gamification in education is to increase student motivation \cite{10.1145/3313831.3376882, 10.1145/2858036.2858231}, engagement \cite{10.1145/3290607.3312806}, and learning performance \cite{10.1145/3173574.3173885}. However, some scholars have cautioned that using gamification is not devoid of potential side effects \cite{10.1145/985692.985741, Andrade2016, Toda2018, bogost2014gamification}. One of the overarching concerns is that the \textbf{gamification itself might become a new source of distraction to learning} \cite{Kim2016, Toda2018, 10.1145/3441000.3441073}. 

\textit{Gamification misuse} occurs when students fixate too much on gamification and get distracted from learning \cite{10.1145/3441000.3441073, 10.1145/3341215.3356335, 10.1145/985692.985741}. Understanding this phenomenon assumes importance for the Human-Computer Interaction (HCI) and Learning at Scale (L@S) research communities because such misusing behavior wastes students' valuable time and negatively affects their learning efficiency \cite{Toda2018}. However, the phenomenon of gamification misuse is still relatively an underexplored research topic in the literature of HCI and L@S. Most of the prior research has only investigated the positive aspects of gamification or whether students have been laboring under gamification or not (i.e., behavioral engagement) \cite{10.1145/2724660.2724665, 10.1145/3173574.3173885, 10.1145/3449086}. Consequently, not much attention is paid to the quality of students' interactions with gamification and their actual learning experience.

\textbf{This paper} aims to fill this knowledge gap about gamification misuse by taking the first steps to study the phenomenon in a large-scale gamified learning app. More specifically, our study is guided by the following three research questions:
\begin{itemize}
    \item \textbf{RQ1:} How do learners perceive and experience the ramifications of gamification misuse? 
    \item \textbf{RQ2:} What factors lead learners to misuse gamification in a learning app?  
    \item \textbf{RQ3:} How can gamification designers and practitioners deal with the problem of gamification misuse?
\end{itemize}

We conduct our research on the world's most downloaded educational app called Duolingo \cite{Duolingo_number_1}. Duolingo is a language learning app that provides over 60 different language courses in more than 20 languages \cite{10.1145/3231848.3231871}. As a case study, Duolingo features an excellent example of gamified learning app for its scale and diversity of game design elements, including but not limited to experience points (XP), XP boosters, leaderboards, badges, streaks, Duolingo currencies (gems and lingots), and interactive stories (see \cite{Shortt2021}). 
 
Due to the explorative and complex nature of our study \cite{Blandford2016}, we take a qualitative approach to answering our research questions. Our study comprises two parts: (1) a large-scale \textit{qualitative content analysis} on Duolingo's user-generated forum data from the past nine years and (2) follow-up \textit{semi-structured interviews} with 15 international Duolingo users. Through a thematic analysis on data from (1) and (2), we present the first documented \textit{taxonomy} for the impact of gamification misuse on users' experiences (from their own perspective). Furthermore, we add nuance and sophistication to the literature by identifying the reasons users tend to misuse gamification and grouping them into two categories: \textbf{active reasons} (i.e., \textit{competitiveness}, \textit{overindulgence in playfulness}, and \textit{challenging the system}) and \textbf{passive reasons} (i.e., \textit{dark nudges of gamification}, \textit{compulsion}, and \textit{herding}). We conclude this research by providing some practical suggestions to gamification designers and practitioners to help them avoid or reduce the instances of gamification misuse in future designs. 

\noindent\textbf{Contributions.} This paper makes three important contributions to the HCI research community in general and L@S research community in particular: \textbf{(I)} we present the first qualitative research on the overlooked problem of gamification misuse in learning; \textbf{(II)} our research adds in-depth insights into the dark side of gamification in learning and thus contributes to the diversity of theories in HCI and L@S; and \textbf{(III)} we introduce some new design and research opportunities for the future use of gamification in learning.

\begin{figure*}[th!]
     \centering
    \begin{minipage}[b]{0.22\textwidth}
         \centering
         \includegraphics[width=3cm, height=5cm]{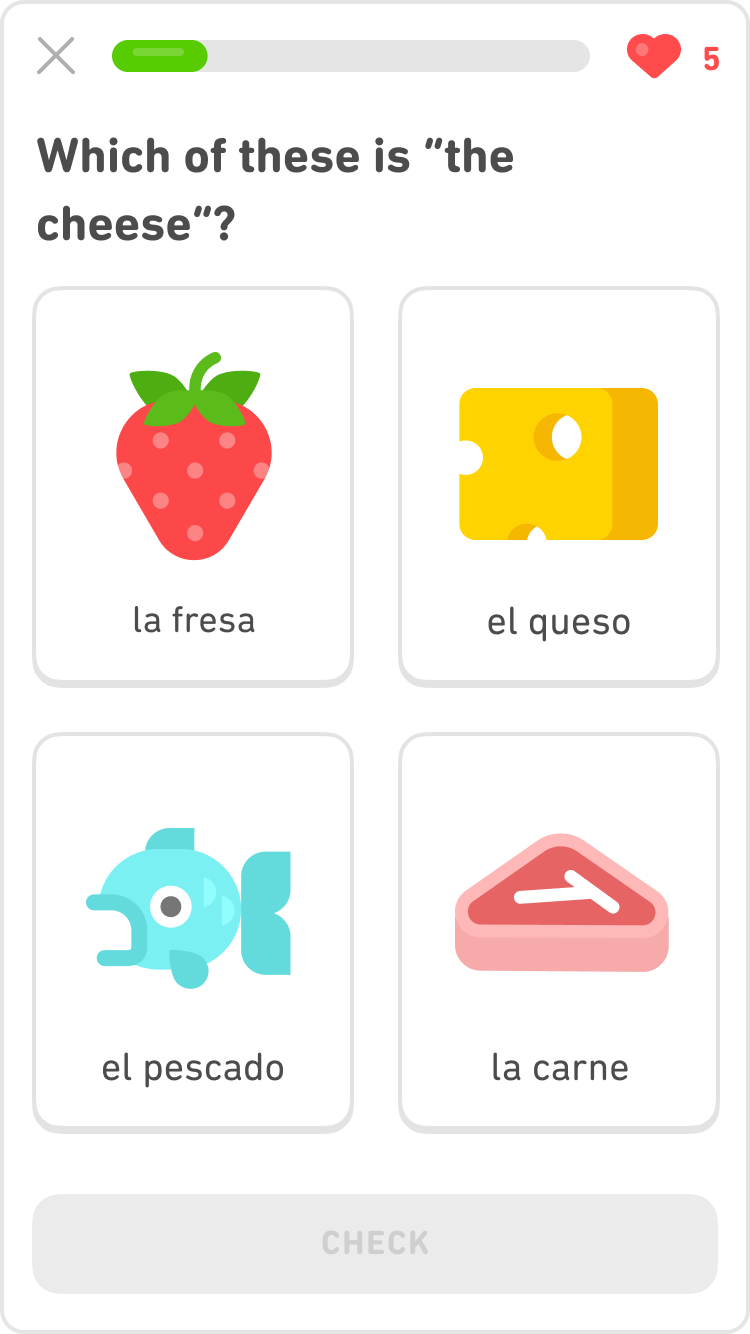}
         \subcaption{A sample question}~\label{fig:question}
     \end{minipage}\hfill
    \begin{minipage}[b]{0.22\textwidth}
         \centering
         \includegraphics[width=3cm, height=5cm]{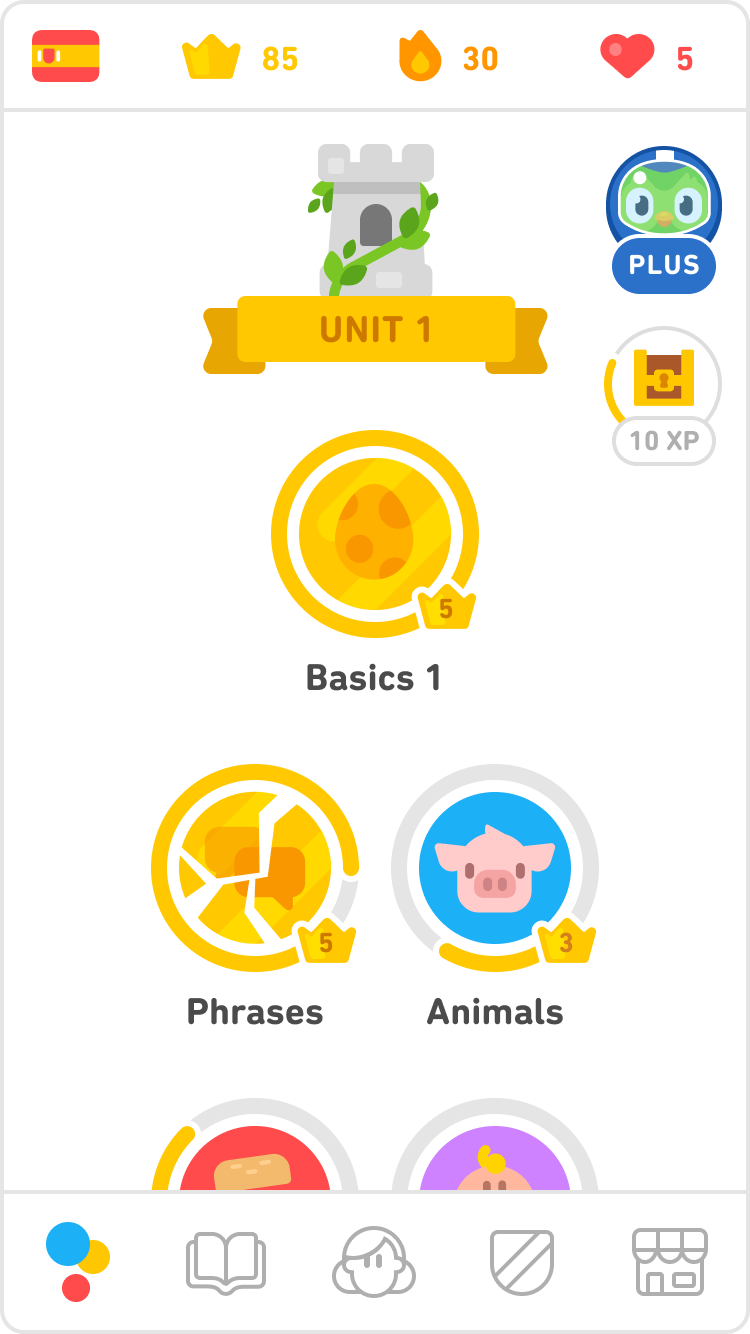}
         \subcaption{The learning tree}~\label{fig:tree_1}
     \end{minipage}\hfill
    \begin{minipage}[b]{0.22\textwidth}
         \centering
         \includegraphics[width=3cm, height=5cm]{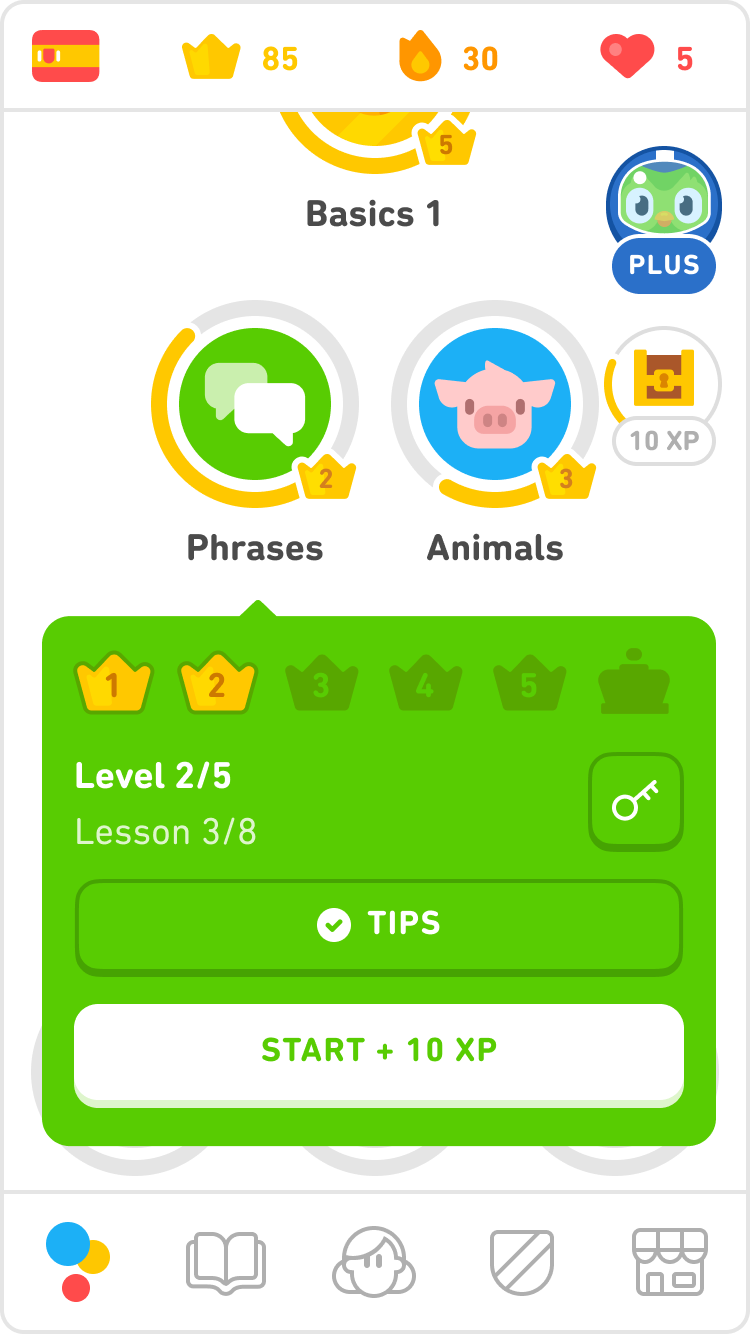}
         \subcaption{Crowns}~\label{fig:crown}
     \end{minipage}\hfill
    \begin{minipage}[b]{0.22\textwidth}
         \centering
         \includegraphics[width=3cm, height=5cm]{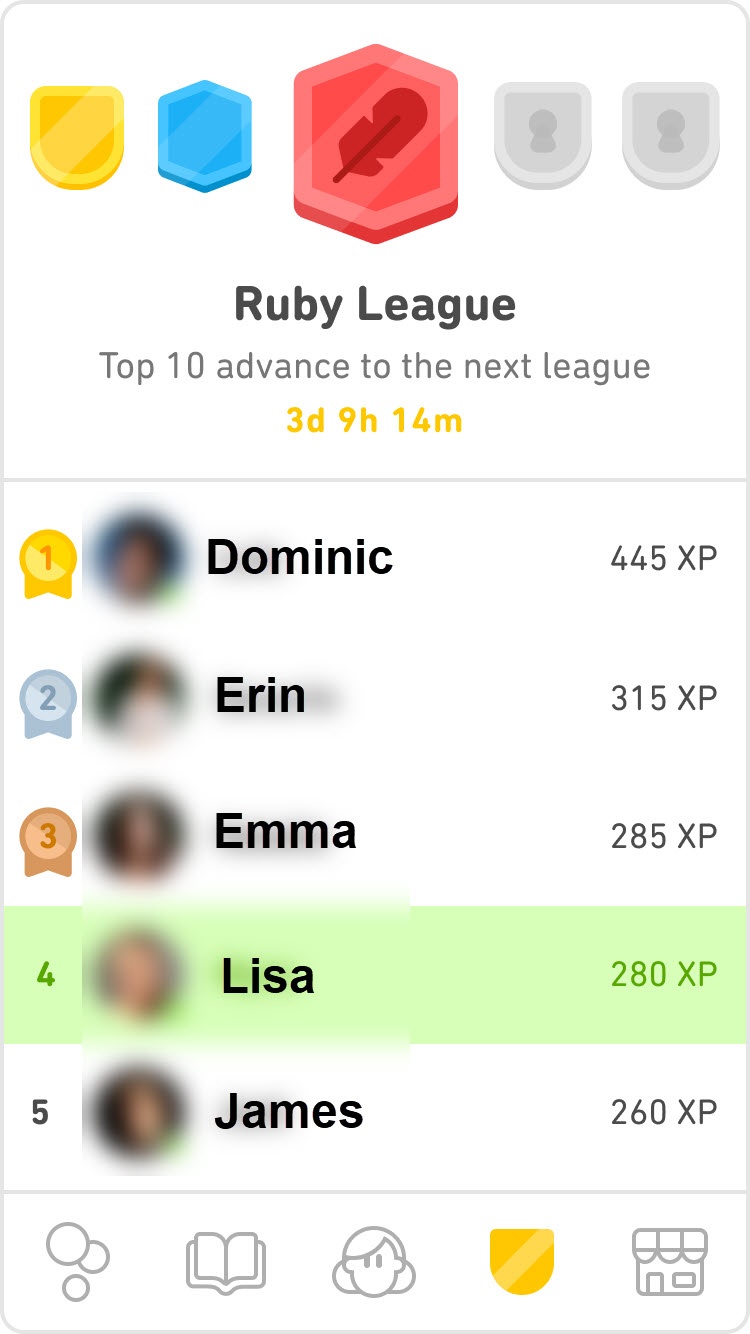}
         \subcaption{A Leaderboard in Ruby}~\label{fig:league_leader}
     \end{minipage}\hfill
        \caption{Anonymized screenshots from the learning application of Duolingo. Figure (a) shows a sample multiple-choice question exercise. Figures (b) to (d) show some variations of gamification rewards in Duolingo. All of the screenshots are credited and provided by Duolingo.}~\label{fig:SnapshotsHDU}
\end{figure*}

\section{Related Work and Background}\label{sec: RW}
\subsection{Gamification, Motivation, and Learning}
One of the most important opportunities for the use of gamification lies in the field of educational technologies \cite{10.1145/2858036.2858231, 10.1145/3383456}. In general, gamification implies using game design elements in non-game contexts \cite{Seaborn2015, 10.1145/2181037.2181040}, and it is typically implemented via the \textit{BLAP system} put forward by Nicholson, which encompasses using Badges, Levels/Leaderboards, Achievements, and Points \cite{Nicholson2014}. The basic aim of these awards is to encourage individuals to perform specific activities and achieve particular accomplishments \cite{10.1145/2488388.2488398, 10.1145/3178876.3186147, 10.1145/2181037.2181040}.

A recent study conducted by the famous Gamification Group of Tampere University shows that using gamification for educational purposes can improve students' performances (in the learning tasks) by up to 89.45\% \cite{Legaki2020}. In another study, researchers have shown that 67.7\% of students prefer gamified learning better than the traditional (so-called passive learning \cite{mintzes2020active}) courses \cite{Chapman2018}. All of these findings point to a promising future for the growth of gamification in education \cite{Swacha2021}.

At its core, gamification aims at enhancing people's \textit{motivation} to perform various tasks \cite{10.1145/2858036.2858231, 10.1145/3311350.3347167, 10.1145/2695664.2695752}. In general, motivation can come from two sources, whether \textit{intrinsic} or \textit{extrinsic} \cite{Jin2021, 10.1145/3313831.3376723, 10.1145/3173574.3174174}. \textit{Intrinsic} motivation occurs when there is no compensation or reward for performing a particular action, and the action itself is perceived as interesting and enjoyable \cite{10.1145/3313831.3376723}. In contrast, \textit{extrinsic} motivation refers to a situation in which people perform activities to achieve a desired outcome other than the activity itself (e.g., to earn more revenue) \cite{10.1145/2858036.2858231, 10.1145/3313831.3376723}. Gamification is thus inherently related to the latter type of motivation \cite{10.1145/2583008.2583029}. 

Furthermore, previous research has shown that motivation is an essential component of education that ultimately affects how students interact with learning platforms or essentially learn \cite{10.1145/2858036.2858231}. For example, intrinsically motivated learners have been shown to complete their studies with greater engagement (or commitment) than extrinsically motivated learners \cite{Tang2018}. 

According to Deci, extrinsic motivation can increase or decrease intrinsic motivation, depending on the nature of the external reward \cite{Deci1971, kizilcec2014encouraging}. For example, positive feedback (as a gamification element \cite{10.1145/3311350.3347167}) has been shown to significantly boost individuals' intrinsic motivation, while other gamification incentives such as competition and performance-based rewards have been widely criticized because they can undermine learners' intrinsic motivation \cite{deci1999meta, kizilcec2014encouraging}. Furthermore, rewarding previously unrewarded tasks could sometimes change the intrinsic motivation of users into an extrinsic one, also called the \textit{over-justification effect} \cite{Lepper1973}.

Therefore, although gamified learning is being lauded for its ability to engage users and help them achieve suitable learning outcomes, the approach is not devoid of concerns. Next, we review the literature to exemplify the most salient instances of gamification misuse and delineate the most important underlying theories.
\subsection{Gamification Misuse and Theories}
One of the major concerns with gamified apps is that their gamification might be used in ways that are not intended \cite{Andrade2016, Diefenbach2019, Toda2018}. Gamification becoming people's only goal for performing tasks rather than their motivation for doing tasks is among the most cited problems \cite{Andrade2016, Diefenbach2019, 10.1145/985692.985741, Toda2018}. In the context of learning environments, for example, students may be tempted to become too fixated on gamification and get distracted from learning \cite{Andrade2016, 10.1145/3441000.3441073}. Some students might even resort to nefarious strategies such as \textit{cheating} to achieve the desired gamification outcomes or just to accumulate more rewards \cite{Andrade2016, 10.1145/3383456, Ibanez2014, 10.1145/2793107.2810293}. 

However, gamification misuse can also take other forms, such as exhibiting \textit{off-task behaviors} \cite{Andrade2016}, \textit{addiction} \cite{zichermann2011gamification, Yamakami2013, Andrade2016}, \textit{the development of a speculative or gambling mindset} \cite{Yamakami2013}, or \textit{obsession with undesirable competitions} \cite{Andrade2016, Yamakami2013}. With these examples in mind, most studies suggest that the loss of intrinsic motivation for learning due to the dominance of extrinsic motivation is the main reason for the misuse of gamification \cite{Featherstone2019, vanRoy2019, Rapp2019, Leito2021}.
 
In the remainder of this subsection, we elaborate on two important underlying theories that can shed further light on why gamification is misused. \textit{Self-Determination Theory} (SDT) is currently one of the most prominent motivational theories in the field of HCI \cite{10.1145/3313831.3376723}. According to this theory, human beings have three innate needs that lead them to behave or take an action the way they do \cite{10.1145/3313831.3376723}: \textit{1) competency} (efficacy), \textit{2) relatedness} (social interaction), and \textit{3) autonomy} (personal agency/volition). Frustration in any of these three basic needs can move people to show misusing behavior \cite{KanatMaymon2015}.

Another established framework for investigating the roots of gamification misuse is known to be the \textit{Rational Choice Theory} (RCT) \cite{10.1145/3235765.3235803, 10.1145/3313831.3376570}. RCT explains that people are naturally inclined to perform actions that appear to be to their advantage. In the case of gamification misuse, what seems advantageous to students is twisted. They consider rational only those actions that help them win rewards and gain immediate mental satisfaction. Thus, they become reluctant to engage with the actual learning. Therefore, as RCT states, gamification misuse becomes more likely when the expected benefit of the misuse (after subtracting the expected disutility of the risk of being caught or punished) is greater than the benefit of the available alternative actions \cite{10.1145/3235765.3235803}.

\textbf{Reflection.} Motivational theories, SDT, and RCT provide interesting viewpoints for why gamification is misused. However, these theories and their subsequent explanations are often too general for improving the gamification schemes in practice. Furthermore, they do not tell us much about the ramifications of misuse. Therefore, our research will help complement these theories by providing more nuanced and in-depth views of gamification misuse.
\subsection{Duolingo: A Brief Introduction}
Launched in 2012, Duolingo is a language-learning app that has since garnered the attention of millions of people who want to learn new languages \cite{10.1145/3099023.3099112, 10.1145/3386527.3406753, Duolingo_website_cite}. Language learners can use the app to engage with different types of learning activities that include studying vocabulary flashcards, doing listening exercises, and solving multiple-choice and shuffled sentence questions. Figure \ref{fig:question} shows an example of a multiple-choice question in Duolingo. The complete list of all exercises in Duolingo can be found in \cite{Duo_ex}.

One of the best-known features of Duolingo is its gamification. As users learn new language skills and brush up on their previous knowledge, they can earn various types of gamification rewards such as experience points (XP), badges, and streaks. Figures \ref{fig:tree_1}, \ref{fig:crown}, and \ref{fig:league_leader} show some of the different variants of gamification rewards in Duolingo. Unversed readers who are not familiar with Duolingo can find detailed information about this application and its features in Appendix \ref{Duolingo_detailed}. This background information is needed to communicate our findings better to a broader audience in the HCI and L@S research communities more effectively.

\section{Research Methods}
This section describes the methodology and data used in our work and consists of three parts: Content Analysis (CA), Semi-Structured Interviews (SSI), and Research Ethics.

\subsection{Content Analysis (CA)}
We employ content analysis because it can classify and summarize the views of a larger number of users for whom access is not otherwise available \cite{nicholas2017reviews, smith1992motivation}. Given that content analysis is viewed as a non-reactive research method, it can be used to unobtrusively understand the most candid and honest opinions of users \cite{smith1992motivation}. Moreover, this phase enables us to establish a foundation for our imminent semi-structured interviews by forming a more comprehensive interview protocol \cite{ma2018professional}. The following subsections provide more information about the data collection, data description, and analysis pipeline.
\subsubsection{Data Collection}
After obtaining permission from Duolingo, we proceed in three steps to crawl the publicly available data (from Duolingo) for our content analysis. \textit{First}, we use the literature (see \cite{Shortt2021, Andrade2016}) to extract 29 keywords related to the themes of `gamification' and `misuse' to use for finding the relevant content from Duolingo forums.\footnote{[`gamification', `league', `leaderboard', `EXP',
`experience point', `badge', `lingot', `tree',
`crown', `streak', `character', `score', `gem',
`heart', `interactive', `point', `power-up',
`reward', `achievement', `misuse', `useless', `harmful', `unwanted', `cheat', `immersive', `leave', `unlock', `forced',
`anomaly']} The keywords are intentionally broad to capture as many potentially relevant posts as possible. \textit{Second}, we use Duolingo's search function, available in Duolingo's discussion forums \cite{Duolingo_disc_search}, to search for each of our keywords. Each keyword returns a maximum of 20 posts. \textit{Finally}, we use \textit{Helium Scraper (v.3)} \cite{Helium_Scraper} to obtain the links of the search results and then access the content of each post. It is worth noting that some user posts have more than 300 attached user comments. These comments are important because they show the subsequent discussions in the community about each post.
\subsubsection{Data Description}
In total, we collect more than 30,000 pieces of text from Duolingo's forum data, covering a period of more than nine years, i.e., from \textit{June 7, 2012} to \textit{July 30, 2021}. We store all data in Microsoft Excel files based
on our corresponding keywords in alphabetical order, following the original order of chat threads sorted by default in each post. According to the statistics at hand, there are 357 posts and 30,618 comments in our dataset, created by 285 and 12,252 users, respectively. Furthermore, the posts and comments have an average word count of 173.76 (SD=66.12) and 27.88 (SD=11.12), correspondingly. Most of the posts (76.47\%) are found from the section \textit{Duolingo and Troubleshooting} of the forums, while the rest (23.53\%) are found from other sections. In addition, our data includes users with different gamification profiles. When we average the number of user \textit{streaks}, \textit{total XP}, and \textit{total crown} feature, we get the values 203.83 (SD=480.46), 54,337.30 (SD=133,497.41) and 465.22 (SD=837.92), respectively. 
\subsubsection{Analysis Pipeline}
We adopt an integrated deductive and inductive approach to carry out our thematic analysis and investigate the qualitative data \cite{lyon2019use}. The deductive part assumes significance in that we treat the related work by Kim and Werbach \cite{Kim2016, kim2015gamification} and Toda \cite{Toda2018} as the basis on which we further develop our inductive content analysis. The inductive part enables us to accommodate the possibility of adding any new themes (or sub-themes) that might directly emerge from the data (similar to \cite{10.1145/3430895.3460126}). 

Our analysis pipeline comprises the following: Firstly, two researchers from the authors (henceforth analysts), select 1,000 random samples each from the dataset to get familiar with the study's content, which signifies a common approach in carrying out qualitative studies where the dataset size is large \cite{sengupta2020learning, li2018working, sullivan2012s}. Secondly, the analysts undertake a detailed assessment of the content and study the entire data from the original un-sampled dataset to develop their initial codes in an independent manner. For this purpose, they utilize a qualitative analysis application called \textit{ATLAS.ti} \cite{Atlas_site} to organize their findings that are then shared and discussed with other team members in weekly group meetings.

Finally, the Fleiss' Kappa measure is employed for estimating and reporting the agreement between each team member participating in the process of coding. Notably, the \textit{reliability analysis tool} available in the \textit{IBM SPSS Statistics} \cite{SPSS_IBM_site} application is used to calculate the measure. Our study reached a Kappa measure of $\kappa=0.87$ indicating a near-perfect agreement between each of our team members on the final themes \cite{cambre2018juxtapeer, 10.1145/2998181.2998342}. 

\begin{table*}[th!]
\centering
\caption[Caption for LOF]
{Demographics of Participants}~\label{tab:demographics}
\includegraphics[width=0.65\textwidth]{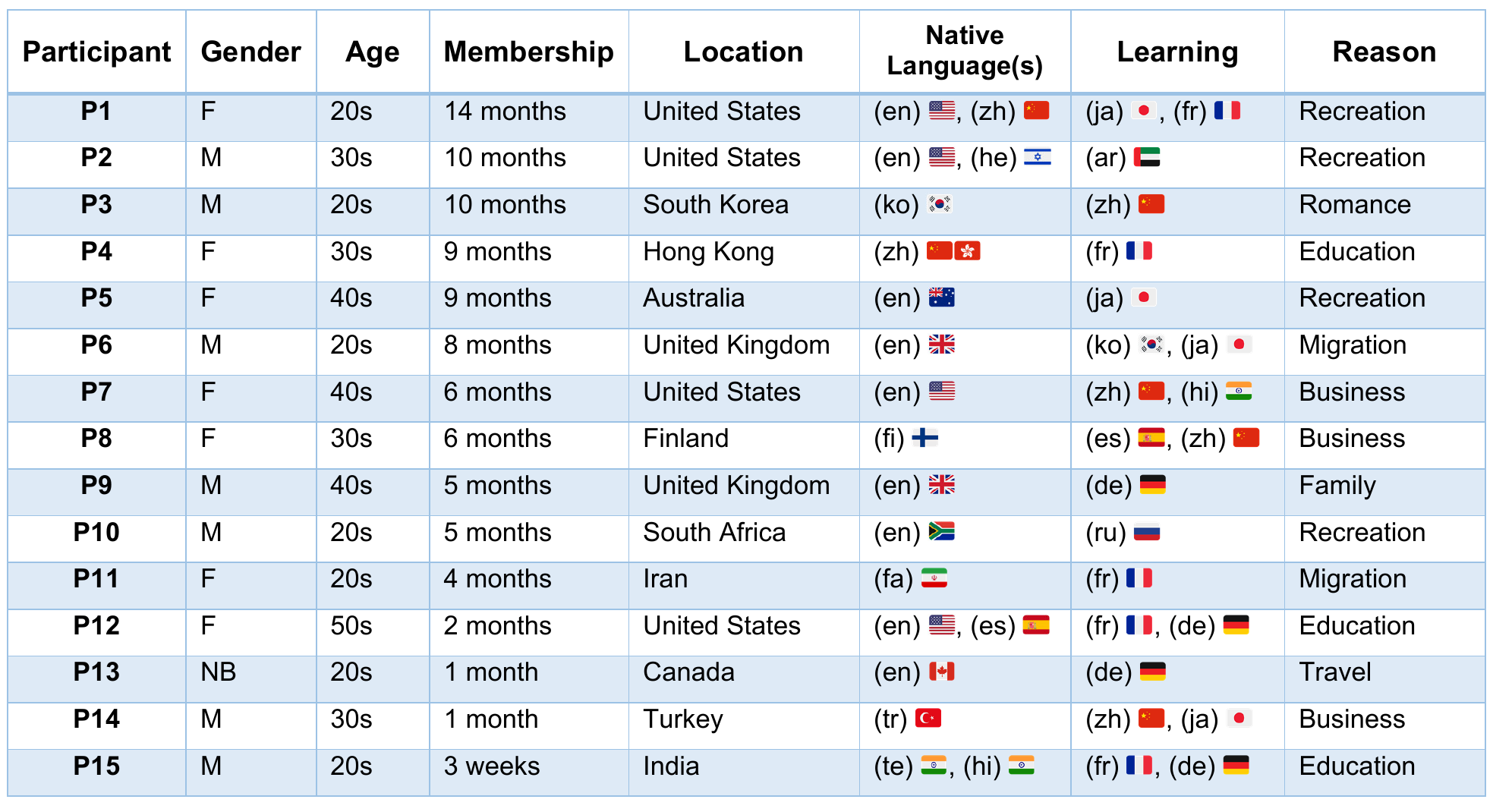}\\
Note: To get a tidier table, we show each language with a two-letter lowercase abbreviation (see ISO 639-1 \cite{iso_standard}). Using flags for showing languages is inspired by Duolingo app itself (see \cite{all_flags, Duolingo_website_cite}).
\end{table*}
\subsection{Semi-Structured Interviews (SSI)}
To triangulate our prior findings and seek any other themes that may have been omitted owing to the restrictions of the text-based content \cite{ma2018professional, ding2020getting, leavy2014oxford}, we complement our study by conducting a series of semi-structured interviews with 15 Duolingo users. In the following subsection, we provide additional information about the recruitment, research participants, interview procedures, and the analysis pipeline.

\subsubsection{Recruitment}
Notably, recruiting participants with the experience of misusing gamification and soliciting their opinions is challenging. First of all, when users do something unruly, they usually prefer to hide it or not talk about it. Second, since the number of complaints about gamification misuse in content analysis is in the minority (around 21\%), it is reasonable to assume that the number of misusers should also be fewer than the regular users. Owing to this, our research combines snowball and convenience sampling methods to recruit its research participants \cite{10.1145/2675133.2675148, 10.1145/2750858.2804281}.

We share digital flyers for recruitment of participants on the discussion forums of Duolingo and other social media platforms such as Facebook, Twitter, and WeChat (similar to \cite{10.1145/3359259, dillahunt2020positive}). In order to be eligible for this study, all participants must (1) be above 18 years of age, (2) fill out and return an informed consent form, (3) have a valid Duolingo account, and (4) have previous experience of gamification misuse in Duolingo.

\subsubsection{Participants} 
Our participants are 15 international Duolingo users (P1-P15: 7 female, 7 male, 1 non-binary) in the age range between 20 and 54 years old (M=29.86, SD=10.90). We recruit them from a wide range of demographic backgrounds from around the world. Table \ref{tab:demographics} shows where each participant is from, what their native language is, and what they are learning on Duolingo. Our participants have different reasons for learning their target languages: \textbf{Recreation} (4), \textbf{Business} (3), \textbf{Education} (3), \textbf{Migration} (2), \textbf{Family} (1), \textbf{Romance} (1), and \textbf{Travel} (1). According to the Common European Framework of Reference for Languages (CEFRL) \cite{language_level}, all of our research participants self-report that they consider themselves to be beginner or elementary level (i.e., levels A1 and A2) in the target languages they are learning. With the exception of P4, P6, and P11, who take serious language learning courses, and P14, who also uses \textit{Memrise}, all other participants rely only on Duolingo to learn their target languages. The membership duration and experience of our participants with Duolingo ranges from 3 weeks to 14 months (M=6.05 months, SD=3.79). When we average the number of user \textit{streaks}, \textit{total XP}, and \textit{total crown} feature, we get the values 155.69 (SD=55.38), 41,118.90 (SD=4,865.30) and 153.12 (SD=49.60), respectively. 
\subsubsection{Interview Procedure} Before commencing our interviews, we provide all interviewees with a copy of our interview questions at least one day prior to the meeting (similar to \cite{10.1145/3430895.3460126}). This approach gives interviewees enough time to familiarize themselves with the interview questions and carefully think about their answers. All the interviews in this work are conducted online using Zoom or Skype. During the interview, two researchers from the authors collaborate to ask questions and take notes. All sessions begin with a succinct introduction of the research and their duration is nearly an hour (M=48.2 minutes, SD=8.71). After seeking our interviewees' consent, we audio-record and transcribe the session for additional analysis. Our interview questions are organized into four distinct themes of \textit{warm-up questions}, \textit{ramifications}, \textit{reasons behind misuse}, and participant's \textit{suggestions}. Upon the end of this study, we email an electronic Amazon gift card worth \$20 to all interviewees as a token of our gratitude.

\subsubsection{Analysis Pipeline}
As far as the interview data is concerned, our analysis pipeline is similar to the approach adopted during content analysis. For this phase, the leading analysts are the researchers who have also held the interviews. After independently coding the data, we discuss the findings in weekly group meetings. It does not take us more than one month to analyze the entire interview data. The Fleiss' Kappa measure of $\kappa=0.89$ in our study denotes a near-perfect consensus between all our team members with respect to the eventual thematic results.

\subsection{Research Ethics}
In general, we ensure compliance with the ethical guidelines of HCI and L@S research to gather the data of users and perform our analyses (see \cite{GDPRref, AoIRref, Bruckman2014}). This is inclusive of seeking approval from Duolingo (for content analysis), interviewees (for interviews), and our university's Institutional Review Board (IRB) (for both) prior to commencing our study. Furthermore, all the analysts involved in this research have obtained certified training in human subjects' protection regulations. All of our interviews and meetings are conducted online to avert any risk of being exposed to the variants of COVID-19.
\begin{figure*}[t!]
     \centering
         \includegraphics[width=0.90\textwidth]{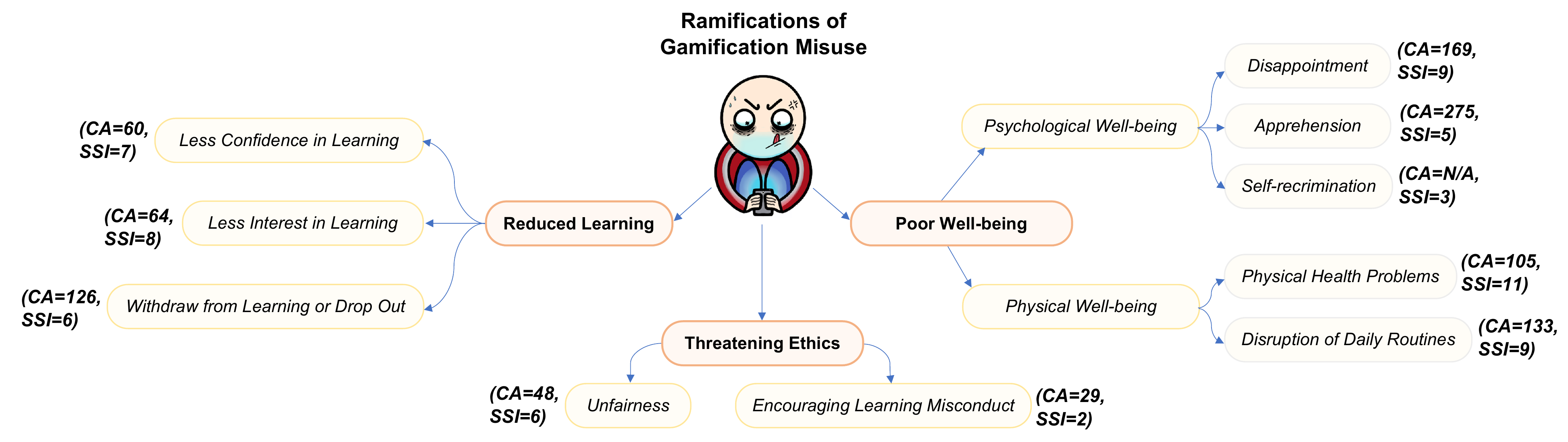}
         \caption{The thematic network for RQ1. We find that gamification misuse can be a threat to user \textit{learning}, \textit{well-being}, and \textit{ethics}. The values of CA and SSI show the number of times each theme has appeared in our content analysis and semi-structured interviews, respectively.}~\label{fig:TN11}
\end{figure*}
\section{Findings}
Based on our research questions, we organize the findings into three main sections: user experience and perceptions (RQ1), uncovering the reasons behind gamification misuse (RQ2), and suggestions for gamification designers (RQ3). The main themes of each section are identified with bullet points ($\bullet$). To make it easy to follow the text, we further itemize the findings in RQ1, RQ2, and RQ3 with the enumerated letters \textbf{E} (\textit{\textbf{E}xperiences}), \textbf{R} (\textit{\textbf{R}easons}), and \textbf{S} (\textit{\textbf{S}uggestions}), respectively. Furthermore, we ``\textit{italicize}'' the responses from both the content analysis and the semi-structured interviews, but only the anonymized identities of the interviewees are attached to their quotes (see Table \ref{tab:demographics}). The numbers after each theme refer to the repetitions of that theme in the quotations from the content analysis (CA) and the semi-structured interviews (SSI).

\subsection{User Experience and Perceptions (RQ1)}~\label{sec:RQ1}
The \underline{summary} of our main findings from RQ1 is as follows.\\ 
\checkmark 1) The misuse of gamification can negatively impact users' learning aptitude and capacity by weakening their confidence in learning. It can also affect users' interest in learning and cause them to give up learning itself or even abandon the learning app.\\ 
\checkmark 2) According to our findings, gamification can sometimes be worrisome concerning users' well-being. The gamification misuse can lead to well-being complications such as disappointment, apprehension, self-recrimination, physical health problems, and disruption of daily life routines.\\
\checkmark 3) Furthermore, the misuse of gamification can deprive users of the right to learn in a fair learning environment. In addition, users believe that the misuse of gamification (if not addressed) can set a bad example for how success is defined in learning communities, allowing this unproductive behavior to become widespread among more users.

\subsubsection{Themes} In the remainder of this subsection, we share an in-depth narrative on the three themes inferred for RQ1 and cite insightful user comments from CA and SSI. Figure \ref{fig:TN11} shows the thematic network \cite{AttrideStirling2001} of our findings from the bird's eye view.\\
$\bullet$ \textbf{E1) Reduced learning} (CA=336, SSI=15). This theme includes adverse effects such as loss of confidence in learning, loss of interest in learning, and dropping out of the learning app.

\textbf{E1-1) Less confidence in learning.} The misuse of gamification reduces the users' confidence in their learning abilities. One of the users explains, ``\textit{[After I stopped the gamification misuse,] it took me over a week to gain [back] the confidence (this is no joke) to [start learning again] (...) I felt like a lead weight had been lifted from me.}'' P14 confirms this point and says: ``\textit{In the end, I completely lost my confidence in truly learning Chinese. (...) Without cheating, most of my answers were wrong because I'd [cheated] before to climb the top rungs of my [learning tree] and gain false [glory] in the leaderboards. (...) I was [enslaved to] the gaming aspects of Duolingo and that heavily shook my confidence in learning.}'' The negative effect proves to be a significant issue because, as it is also mentioned in the literature, users who are more confident in their learning ability tend to show more success in finishing their learning tasks \cite{10.1145/3027385.3027388}.

\textbf{E1-2) Less interest in learning.} We infer that the misuse of gamification reduces the users' interests in learning. This problem is evident in the comments of users who criticize Duolingo for adding new lessons to their learning plan. These agitated users argue that new lessons keep them from instantly achieving all the rewards of gamification. However, the other members, who are not so obsessed with gamification usually confront them by asking, ``\textit{Do you know why you started [using] Duolingo [in the first place]? Yes, because you want to learn a new language (...) [Not] re-gilding your [learning] tree over and over again.}'' We think that P1's viewpoint presents the flavor of the perspective the best: ``\textit{Through the many nights of XP farming, I lost my drive to learn [new languages]. (...) After abusing gamification, you should, of course, expect learning [not to be as] fun as it used to be. [Some days] I was so obsessed that I was literally seeing real learning as an obstacle to my success in gamification.}''

\textbf{E1-3) Withdrawal from learning or dropping out.} Some users opine that the misuse of gamification can increase the tendency of individuals to withdraw from learning and that this withdrawal can even lead them to leave the platform forever or for a long time (known as \textit{Churning} \cite{Mogavi2019, 10.1145/3449086}). One user comments: ``\textit{[Gamification misuse] was my reason to stop two years ago. It felt like I was trapped on Duolingo [and] the fun was gone.}'' According to our findings, it is also noted that when users become obsessively attached to their gamification achievements, it becomes increasingly unbearable for them to continue learning without the gamification accomplishments. One user gives an example for obsessive attachment to streak as an example: ``\textit{My brother lost his 110-day streak, and now [he] is an abandoned account.}'' P2 confirms this point, adding that a false sense of purpose created by the misuse of gamification can lead users off their learning path and ``\textit{even force them to give up language learning of any kind...}'' P2 continues, ``\textit{I think that the destructive effect [could] be [even] worse if [the user] is a child or a teenager.}'' Over time, this problem can also jeopardize the sustainability of the learning application \cite{10.1145/3449086}.\\
\noindent$\bullet$ \textbf{E2) Poor well-being} (CA=487, SSI=11). According to the literature, well-being is an umbrella term that refers to an individual's life satisfaction, positive emotions, and lack of stress or physical tension \cite{10.1145/3290607.3298998, 10.1145/3290607.3312901, 10.1145/3411764.3445771, 10.1145/3313831.3376176}. In the context of this definition, our emergent second theme focuses on the negative impacts of gamification misuse on user well-being. According to our findings, most concerns are related to the impact of gamification misuse on users' psychological well-being. However, there are some physical complications reported as well.

\textbf{E2-1) Disappointment.} First and foremost, the users talk about their feeling of disappointment after misusing gamification. Disappointment is a psychological reaction that occurs when an outcome does not meet an individual's expectations \cite{vanDijk1997, chauveau2009theory}. Users are either not satisfied with their performance or fail to establish an achievement they were looking forward to. P8 sums up the general view of this sub-theme the best and explains, ``\textit{It is more disappointing to lose even after ... cheating ... [and it happens] because there are always more and better cheaters out there. The bitter truth is that no one can remain number one in the Diamond league rankings forever, and thus gamification abuse is I think a bottomless pit. At some point, the abuser should come to his senses and acknowledge that Duolingo's gamification is always the final winner (...) I wish I [were] more focused on learning and had not wasted my time over nothing.}'' 

\textbf{E2-2) Apprehension.} Another adverse effect of gamification misuse on psychological well-being, according to users, is apprehension. Apprehension is defined as discomfort, stress, or anxiety about an upcoming situation or event \cite{psychology_ref, psychology_ref2}. A new user who seems to be obsessed with the gamification aspects of Duolingo mentions, ``\textit{It is very difficult to not get stressed over the leagues and worry about if I will advance or not instead of focusing on what's the purpose of Duolingo, which is learning languages.}'' Another dissatisfied user confirms this claim and says that gamification on Duolingo is becoming like ``\textit{another [chore] that you have to worry about!}''

\textbf{E2-3) Self-recrimination.} Another problem that is extracted exclusively from the interviews is self-recrimination for gamification misuse, which usually emerges after disappointment and apprehension. In this state, users feel excessively guilty for their obsessions or abusive actions, which self-reportedly has a negative impact on their well-being. P10 who has probably the most representative opinion on this issue says: ``\textit{I felt guilty about my abuse, and it wasn't just a normal sadness. (...) This mood was killing my motivation for learning and [left me] paralyzed for weeks, eating me up from within. (...) So it's wise to start learning with the right [intentions] from the beginning.}'' 

We should mention here that we have not counted healthy sorrow or users' reflections on their past actions as self-recrimination, e.g., ``\textit{Because of my competitive nature, I cheated, knowing it was wrong, just to win. I feel ashamed of my self.}'' Research has shown that such pensive feeling (if moderate) is healthy for the user's conscience and helps prevent future misconduct \cite{Gray2012}. Thus, self-recrimination that is intended in our work encodes only the extreme feelings of guilt where users clearly state how their well-being is compromised.

\textbf{E2-4) Physical health problems.} In addition to these psychological well-being problems, quotes about physical well-being problems are also prevalent in user-generated content and interviews. For example, one troubled user mentions, ``\textit{I was in a quandary. I wanted to learn; not compete! I [would find] myself practicing until 3 a.m. after 6 or 7 hours because I fell into the competitive trap. I was suffering from repetitive strain in my wrist and shoulder. Crazy!}''  We believe that this information is significant because the existing literature has primarily focused only on the negative consequences of video games on well-being and physical health (see \cite{Hussain2011, Lee2016, Chen2016}). As such, our work is the first to highlight the similar negative consequences of gamification.

\textbf{E2-5) Disruption of daily life routines.} Furthermore, in the category of physical well-being issues, users also talk about the resultant disruption to their daily life routines. For example, another user who agrees with the negative effects of gamification misuse on the user's physical health explains how his wife was angry at his obsessive addiction to gamification and consequently the ``\textit{negligence of more important daily life tasks like clearing up the kitchen, making up the bed, and fixing the vacuum cleaner.}''

\noindent$\bullet$ \textbf{E3) Threatening ethics} (CA=61, SSI=6). This theme summarizes users' views on how gamification misuse threatens ethics. 

\textbf{E3-1) Unfairness.} Most users believe that some cases of gamification misuse like cheating can make the learning environment unfair (and even intolerable) to fair-playing users who are truly determined to learn something from the app. One user, who is annoyed by unfair practices and cheaters, explains, ``\textit{[Cheaters] really [take] away any motivation [for learning]. (...) I'm just chasing cheaters up the leaderboards.}'' In confirmation, P14 says, ``\textit{[Gamification misuse] is unethical because it wastes everyone's precious time... probably violates [Duolingo's] terms of service, and, if it is something like cheating, gets on the nerves of classmate who have made an honest effort to win.}'' 

\textbf{E3-2) Encouraging learning misconduct.} Some other users are concerned about the detrimental influence of the rife of gamification misuse on new users and youngsters. A concerned user explains, ``\textit{Duolingo turns many people into worse versions of themselves, and it's a very irresponsible thing, especially on a learning platform like this one. Many kids are here, and Duolingo is actually teaching them how to get greedy, how to manipulate, how to cheat, how to fight for meaningless and utterly useless virtual gems. Yes, they might learn a few words on the way, but their brains will also pick very bad habits and patterns.}'' P9's account on this matter is also similar: ``\textit{I think that every fraud that's tolerated encourages more fraud... and spreads like a virus.}''

\begin{figure*}[th!]
     \centering
         \includegraphics[width=0.90\textwidth]{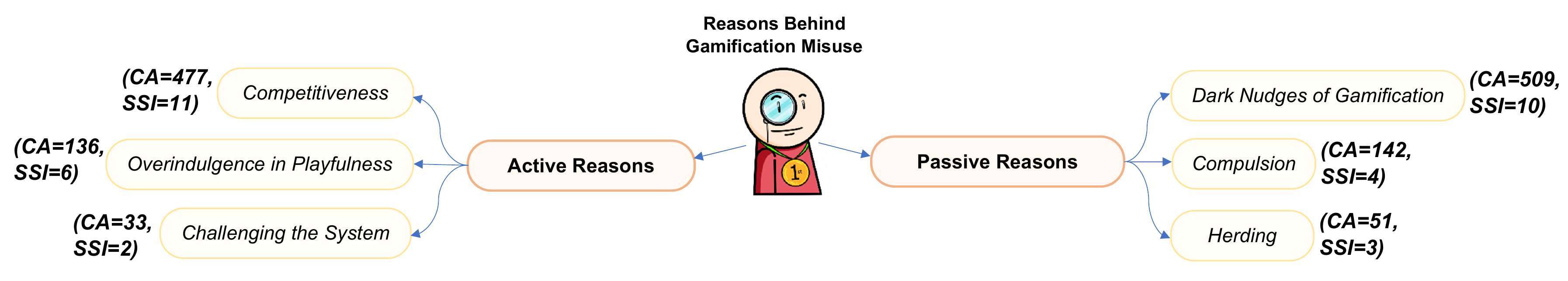}
         \caption{The thematic network for RQ2 provides the detailed reasons users misuse gamification in learning. The values of CA and SSI show the number of times each theme has appeared in our content analysis and semi-structured interviews, respectively.}~\label{fig:TN22}
\end{figure*}
\subsection{Reasons Behind Gamification Misuse (RQ2)}
We find and classify the main reasons for the misuse of gamification into two categories: \textit{active} and \textit{passive}. The \underline{summary} of the main findings from RQ2 is as follows.\\
\checkmark We define the \textit{active reasons} for gamification misuse as those reasons for which users acknowledge their own agency, control, or direct responsibility. We identify three active reasons for the misuse of gamification: \textit{competitiveness}, \textit{overindulgence in playfulness}, and \textit{challenging the system}.\\
\checkmark We define the \textit{passive reasons} as those reasons that indirectly manipulate the user's behavior to misuse gamification. In some cases, users are not even aware of their misusing behavior until they notice it later. We identify three passive reasons for the misuse of gamification: \textit{dark nudges of gamification}, \textit{compulsion}, and \textit{herding}.

\subsubsection{Themes} In the remainder of this subsection, we will take a closer look at the active and passive reasons for misusing gamification. Since different active and passive reasons could also combine and reinforce each other, some user quotes emerge to be naturally multi-faceted. Figure \ref{fig:TN22} shows the thematic network of our findings.\\  
$\bullet$ \textbf{R1) Active reasons} (CA=606, SSI=13). Users might actively misuse gamification for several reasons. 

\textbf{R1-1) Competitiveness.} The first and most common reason cited by users is satisfying their competitive nature. According to the literature, competitiveness is not always an evil quality; the desire to compete with others can be traced to people's inherent tendency to progress or survive in life \cite{cooke2010effects}. Nevertheless, our findings suggest that users who are obsessed with it (especially in a learning environment) can become aggressive and stray from learning. For example, one user comments, ``\textit{I initially was spurred by the capability to `outrank' other users, and I made sure to climb the [leaderboard] until I reached Ruby League. (...) The entire time I was striving to rank up, I didn't even care about what I was learning. (...) The only thing I cared about was beating out the other users.}'' P10 also confirms this point and adds, ``\textit{I resorted to abusing gamification... with cheating and stuff... to prove to everyone who's the boss.}'' 

\textbf{R1-2) Overindulgence in playfulness.} The second most frequently cited reason is overindulgence in playfulness. Users who mention this reason often confuse Duolingo with pure video games such as Candy Crush \cite{candy_crush} or Angry Birds \cite{Angr_Birds}, and \textbf{expect exactly the same joyful effects}. One user mentions, ``\textit{We all are here to play a game though. Duolingo has never been the most efficient or quickest way to learn a language. If someone wants to learn a language fast and thoroughly, it would be best to take some professional courses.}'' However, this kind of attitude towards gamification annoys most fellow learners, and they complain: ``\textit{For users not here primarily to learn a language, why not play Angry Birds instead?}'' A typical response these users hear is, e.g., ``\textit{I [am doing] this (gamification misuse) for fun. It's not like you'll get fluent using Duolingo. For me, Duolingo is just some extra exercise, as I'm already learning my [target] language using [another] program.}'' 

\textbf{R1-3) Challenging the learning system.} Finally, several users mention their desire to challenge the learning system. According to our findings, this could be due to a previous negative experience with the learning app, finding bugs and errors, or simply gaining popularity or monetary gain. For example, P13 says, ``\textit{After Duolingo changed its XP policy, I got mad. I wrote several emails to Duolingo but didn't get a response. So, I was angry and retaliated by gaming the XP system through a paid hacking tool. I thought it was a peaceful way to show my protest, but later I understood that my abuse was hurting the other learners more ... I was completely in my delinquency mood and I am [now] ashamed.}''

\noindent$\bullet$ \textbf{R2) Passive reasons} (CA=672, SSI=12). Sometimes users do not want to misuse gamification on purpose, but some \textit{passive reasons} provoke them to do so and distract them from learning. 

\textbf{R2-1) Dark nudges of gamification.} The first and most frequently mentioned reason in this category is what we characterize as the \textbf{dark} nudges of gamification. In general, nudges are psychologically informed techniques used to steer the user's decisions in welfare-enhancing directions \cite{AlMarshedi2016}. \textbf{Dark nudges} (a.k.a., sludge) are also essentially \textit{nudges}, but with \textbf{harmful, unhealthy, or unproductive purposes} \cite{xiao2021gaming, Newall2018, PETTICREW2020}. The term is relatively new and comes originally from the literature on gambling, where people are enticed to play more games and spend more money on them \cite{Newall2018}. 

We find that some users charge Duolingo of using dark nudges to drive traffic to their platform without caring much about the users' learning experience. The majority of complaints relate to two gamification elements: \textit{leagues} and \textit{leaderboards}. The dark nudges that users have experienced the most are as follows.

$\triangleright$ \textit{R2-1-1) Punishing active or motivated users.} Unfortunately, the current design of the leagues in Duolingo is such that even users who have achieved all of their daily learning goals can be penalized. One user explains, ``\textit{It's so disheartening [to be] demoted from the Diamond league because you `only' [scored] 800 points that week, especially when a daily goal of 50XP is considered `extreme' (according to Duolingo's own daily goal suggestions).}'' This makes some users think that ``\textit{there are more rewards for `playing' the app, rather than learning a language.}'' Several users suspect that ``\textit{[such] design is intentionally exploiting users' psychological resistance against losing the gamification benefits ... for bringing more traffic to the platform;'' ``[otherwise], punishing a motivated user is not reasonable (P11).}'' These users believe that the learning apps must be held accountable for any gamification misuse that arises from this type of paradoxical design.

$\triangleright$ \textit{R2-1-2) Random assignment of users to the competition groups.} In Duolingo, leaderboards are local and group-based (i.e., not global), and they reset on a weekly basis. Each week, Duolingo randomly groups 30 users into a local leaderboard to compete against each other. However, these users might have different learning goals, target languages (to learn), language backgrounds, and study levels. For example, one user might be learning the Japanese alphabet while another is busy with French grammar. As a result, many users find the ranking system in Duolingo's leaderboards inherently unfair and biased. Several Duolingo users argue that this entrenched unfairness and bias may be the reason for some cases of gamification misuse. For example, P9 explains that the ``\textit{human brain would naturally look for a way to resolve the imposed unfairness one way or another, and [users] might be tempted to misuse gamification.}''

\textbf{R2-2) Compulsion.} The next most cited passive reason for gamification misuse is compulsion, which can be defined as a very strong feeling of wanting to do or repeat something that is difficult to control (see \cite{Heather2017}). Research has shown that all people generally experience some degree of compulsion in their daily lives \cite{nice2013obsessive}. However, for some people, compulsion might ensue more severe adverse effects, especially on their daily lives \cite{10.1145/3290605.3300542, 10.1145/1979742.1979750}. For example, in the case of Duolingo, some users report that their compulsion for collecting gamification appellations or rewards (e.g., becoming a ranker in leaderboards, ascension to certain leagues, or collecting particular badges) has totally disrupted their learning experience. One user explains, ``\textit{I often happen to do something I call `binge learning': I do countless lessons just for the sake of doing them and not taking notes, memorizing words, grammar, or anything. It's just so satisfying doing lessons that I can't stop. I know it's a very bad thing to do and harmful to the learning of a language, that requires patience and a slow but steady pace, [but I just can't resist].}'' 

\textbf{R2-3) Herding.} The last most frequently mentioned reason for gamification misuse in this category is herding. It refers to a well-known psychological phenomenon in which ``everyone does what everyone else is doing, even when their private information suggests doing something quite different \cite{10.1145/3433148.3433154}.'' In our work, the term is used to code the scenarios where users use the prevalence of gamification misuse in their leagues or leaderboards as a reason or an excuse for justifying their own misusing behavior. P7's quote probably represents this view the best: ``\textit{I was stuck in a leaderboard where everyone had ten times more points than me. I was one hundred percent sure they were all cheating. No doubt about it. So, I decided to cheat just once ... just to make sure that I wasn't the one who was going to say goodbye to the Diamond League. Our leaderboard had become [the leaderboard of] seeing who can [cheat better], and it was both sad and [funny] ... I know it's wrong, but I'd no choice, and after a point in time, it went out of my control ... They pushed me, and I [became] a fool.}''
\subsection{Design Suggestions for Gamification (RQ3)}
This subsection provides some design suggestions for gamification designers and practitioners working on learning applications to help them overcome or mitigate the problem of gamification misuse. The suggestions extracted here are divided into four distinct themes: (1) personalization and customization, (2) revision of gamification mechanics and dynamics, (3) intelligent detection of gamification misuse, and (4) informing and guiding users.

\subsubsection{Themes} In the remainder of this subsection, we will expound on the four themes extracted for RQ3.

\noindent$\bullet$ \textbf{S1) Personalization and customization} (CA=254, SSI=15). Tailoring of gamification features to adapt users is one of the recurring themes emerging from our analysis. This can be accomplished via \textit{personalization} and \textit{customization} or a combination of the two \cite{10.1145/3290605.3300380}. \textit{Personalization} refers to the process by which a system (e.g., a learning platform) adapts itself (automatically) to the needs and interests of its users. \textit{Customization} is the process that puts the control in the hands of the users, allowing them to modify the system themselves.

\textbf{S1-1) Leaderboards should consider users' compatibility.} Some users (CA =107, SSI =13) contend that all users should have a fair chance to reach the top rankings of the leaderboards. This requires the use of fairer measures for comparing learners. For example, an unsatisfied user has commented, ``\textit{I think it's pretty unfair for someone who is at level 23 in Spanish [lessons] to be competing with someone at level 5.}'' The extant literature of gamification also states that leaderboards can be less motivating (or even off-putting) to users if they do not give all users a fair chance to compete or become successful in the gamified aspects of the app \cite{10.1145/3290605.3300397}.

\textbf{S1-2) Make the gamification schemes optional.} We suggest that users should be able to choose whether or not to use a gamification scheme by their own volition. For example, many users (CA=157, SSI=15) ask, ``\textit{Could there be an option to disable [gamification]?}'' They are of the view that ``\textit{Some, [game-like features] like achievements and leagues, ... can safely be disabled without impacting learning.}'' According to the HCI literature, an optional gamification scheme could also help users' enjoyment and enable them to perform their tasks better  \cite{10.1145/3313831.3376360, 10.1145/3290605.3300380}.

\textbf{S1-3) The diversity of learning scenarios should also be considered.} According to some users (CA=92, SSI=3), different learning scenarios require different designs of gamification. A \textit{learning scenario} is often characterized by variables such as the study session's duration, frequency, and intensity, as well as the learner's competency for learning (see \cite{10.1145/3449086, 10.1145/3290605.3300864, 10.1145/2793107.2793128}). P3 remarks, ``\textit{[Unfortunately], I can only use Duolingo on weekends, and the [daily] streak function is completely unusable for me. (...) The worst thing is when Duolingo wants to warn me on Monday evenings that my [daily] streak will soon be lost. This kind of streak is definitely not my thing. Maybe a weekly streak [would] be better for me. (...) I can well imagine that such an [incompetent] gamification design would tempt people to cheat on the streaks, and they do.}''

\noindent$\bullet$ \textbf{S2) Revising of gamification} (CA=275, SSI=12). Some user suggestions do not fit into the category of personalization and customization. Instead, they ask Duolingo to revise some of its gamification \textit{mechanics} and \textit{dynamics} in general. In the literature, the term \textit{mechanics} refers to functional components of a gamified application (e.g., scoring systems, leaderboards, and levels) that enable various user interactions, controls, or experiences \cite{hunicke2004mda, zichermann2011gamification, thiebes2014gamifying}. The term \textit{dynamics} denotes the run-time behavior of the mechanics, that is, how they affect the user inputs or each other's outputs over time \cite{hunicke2004mda}.

\textbf{S2-1) Difficulty and importance of different learning tasks should be considered.} Some users (CA=170, SSI=9) suggest that gamification rewards should not be given equally to every different tasks at hand. Instead, they should be distributed more wisely based on the \textit{difficulty} or \textit{importance} of the learning tasks. Otherwise, the gamification rewards could lead to a bias against certain learning activities, or even worse, attract more misusers to the learning app by creating unwanted ``\textit{loopholes}.'' One irritated user explains, ``\textit{I am really conflicted about repeating stories. This is one of the cheapest ways to rack up XP if you want to cheat your way to the top of the Leagues.}''

\textbf{S2-2) Learning apps should plan for more than just the behavioral engagement.} The current gamification schemes in learning apps primarily measure and award users' behavioral engagement \cite{10.1145/3449086}. However, some users (CA=164, SSI=7) suggest that cognitive and emotional engagement of users should also be taken into account. One user explains, ``\textit{The current gamification notifications in Duolingo assume that you basically hate learning and you're just in it to compete in leaderboards. That assumption really destroys my inner passion for learning and should change. (P5)}'' Another user affirms and says, ``\textit{I wish the Duolingo system was better able to reward language learning rather than [insane] XP accumulation}.''

\textbf{S2-3) Gamification should encourage users to progress.} To this end, gamification rewards and dynamics should be updated from time to time to avoid learning stagnation. Some users (CA=81, SSI=4) believe that there should always be a stronger gamification-based motivation for users to encourage them to move forward. These updates may comprise decreasing the value of XP for old (and probably easier) lessons over time or increasing the value of the prizes that can be achieved after acquiring more advanced lessons. One user contends that ``\textit{farming the very first skill over and over, using timed practice should not award you always the same amount of XP, because it is promoting cheating!}''

\noindent$\bullet$ \textbf{S3) Intelligent detection of gamification misuse} (CA=97, SSI=10). Despite the dramatic advancements in machine learning and deep learning techniques in recent years, the automatic detection of gamification misuse using such computational models remains an onerous task. The complexity is primarily attributed to the need for copious amounts of labeled (training) data to identify different instances of gamification misuse \cite{10.1145/3290607.3313054, 10.1145/3334480.3375147}. However, including users in the loop to find and label the instances of gamification misuse can be a viable solution. To this end, it might be advantageous to set up a responsive reporting system to collect and track the instances of gamification misuse in a crowdsourcing manner. In this regard, P4 remarks: ``\textit{[Currently,] there's no straightforward way to report gamification abuse in leaderboards ... we can only press the block button ... and that only turns off our means of communication [with misusers] ... and the cheater is still in front of [us] in the leaderboard.}''

\noindent$\bullet$ \textbf{S4) Informing and guiding Users} (CA=N/A, SSI=9). Several users opine that transparent communication with the end-users of gamification might also enable them to reduce their proclivity to misuse gamification.

\textbf{S4-1) Be reminded that not every user knows why misusing is bad.} For some users, the misuse of gamification is pre-defined as an ingenious way to enjoy themselves. Therefore, six users from our interviewees suggest that Duolingo and similar learning apps should confront this type of misuse (due to lack of information) by informing users about the negative ramifications of gamification misuse in advance. Communicating the positive effects of using gamification correctly can also be helpful. P8, who seems to be an ardent supporter of such transparency, notes, ``\textit{When people start to misuse gamification, they [may] be unaware of where it's going ... It may sound like harmless fun to them, but it can cost them their learning and ... perhaps their [mental] health.}''

\textbf{S4-2) Be reminded that some users do not know what gamification is.} Designers and practitioners should have this in their minds that some users do not know what gamification is. Three respondents to our interviews are of the view that the learning apps should inform users in advance about the difference between \textit{games} and \textit{gamification} (refer to \cite{10.1145/2181037.2181040}). According to these users, for some learners (especially the youngsters), it is also important to know why their learning experiences are gamified: ``\textit{leisure, business, game, fun} (P10)'' or ``\textit{an additional support for learning.} (P9)'' Communicating such information might help misusers to use gamification more correctly and responsibly. 

\section{Limitations and Future Work}
In this work, we took the first steps to fill the knowledge gap about gamification misuse in learning apps. Our work contains \textbf{three important take-home messages} that might be insightful for practitioners working in this field: 1) This paper makes clear why gamification misuse and its negative consequences for learning, well-being, and ethics should not be ignored. 2) Our work reveals that all instances of gamification misuse are not intentional. 3) Finally, most user suggestions in our work show that gamification misuse becomes more likely when (gamification) designers do not know their target users very well or have incorrect assumptions about them.

Although our study provides a thorough exploration of users' views on the misuse of gamification, it also poses some drawbacks. Our research is primarily affected by the typical problems associated with qualitative research studies, such as the limited availability of research participants \cite{10.1145/3411764.3445609, 10.1145/2858036.2858395}, the unintended bias of research analysts \cite{10.1145/3313831.3376768}, or unsubstantiated generalization \cite{10.1145/3410404.3414259, 10.1145/3290605.3300698}. 

Furthermore, similar to many other qualitative studies (see \cite{10.1145/3411764.3445609, 10.1145/3411764.3445605, 10.1145/3173574.3174241}), our study explores a qualitative (conceptual) phenomenon (i.e., gamification misuse). The main weakness we would like to point out is that identifying this phenomenon largely depends on whether the end-users perceive it. Therefore, as a next step, we look forward to seeing quantitatively oriented research that helps formulate the definition of gamification misuse and bring it to a (precise) measurable scale that can also be helpful for practitioners. Such quantitative work can be a valuable asset to the L@S community, as it can reduce misinterpretation, improve communication, and avoid redundancy in the field.


Finally, future work may be interested in uncovering the relationship between the negative effects of gamification misuse (Figure \ref{fig:TN11}) and the micro reasons for misuse (Figure \ref{fig:TN22}). We opine that it is a legitimate and important question to ask which reasons are directly responsible for the different effects. Unfortunately, we are unable to provide a conclusive and reliable answer to this question due to the qualitative nature of our research in this study. Hence, as a future work opportunity, conducting a quantitative analysis with hired research participants (on a large scale) would likely be more appropriate to answer this question.

\section{Conclusion}
In this paper, we looked at the problem of gamification misuse in Duolingo, the world’s most popular learning application. We studied the misusing behavior from the users' perspective from three aspects: the negative effects they experienced (RQ1), their reasons for this behavior (RQ2), and finally, their suggestions for practitioners and designers (RQ3). In RQ1, we found that the misuse of gamification could negatively influence users' learning aptitude and capacity, well-being, and even threaten their ethics. In RQ2, we observed several potential reasons for the misuse of gamification. We carefully coded the reasons into \textit{active} and \textit{passive} reasons based on the users' perceived role (agency) in the misuse. In RQ3, we offered gamification designers and practitioners working on learning apps some suggestions to prevent or mitigate the misuse of gamification. All in all, we believe that this paper could serve as an important reference in the HCI and L@S communities for understanding the problem of gamification misuse in learning apps.
\begin{acks}
We would like to thank Duolingo and all of our research participants who made this study possible. In addition, we thank Dr. Lik-Hang Lee and our anonymous reviewers for their valuable suggestions that helped improve the quality of this work. A special thanks also goes to Mr. Rahman Hadi Mogavi for the artistic contributions to this work (i.e., Figures \ref{fig:TN11} and \ref{fig:TN22}). This research has been supported in part by project 16214817 from the Research Grants Council of Hong Kong, and the 5GEAR and FIT projects from Academy of Finland. It is also partially supported by funding from the Theme-based Research Scheme of the Hong Kong Research Grants Council, grant no. T44-707/16-N.
\end{acks}

\bibliographystyle{ACM-Reference-Format}
\bibliography{sample-base}

\appendix
\section{Duolingo: Supplementary Information}\label{Duolingo_detailed}
This appendix provides supplementary information about Duolingo and some of its features. Launched in 2012, Duolingo is a self-paced and free language-learning app co-founded by Luis Von Ahn and Severin Hacker. The application has since attracted the attention of many users around the world who want to learn new languages. English, Spanish, and French are the most popular languages on Duolingo. However, the app also offers minority languages such as Hawaiian, Scottish Gaelic, Navajo, and Yiddish. Users can access various learning materials through Duolingo's official website \cite{Duolingo_website_cite} and mobile apps for iOS and Android. According to Duolingo's official website, the app's primary goal is to make education \textit{free}, \textit{fun}, and \textit{accessible to all}. Two notable features that set Duolingo apart from other language-learning apps on the market are gamification and active discussion forums.

The first feature is \textbf{gamification}. Duolingo is famous for its extensive and varied use of gamification. The app integrates gamification elements such as experience points (XP), badges, streak counts (streaks), and leaderboards into its study plans to make them more engaging and fun for language learners. Each learner's account has a personal learning \textit{dashboard} that tracks her or his XP points and other gamification achievements over time. These learning dashboards have a special gamified construct commonly known as a ``learning tree.'' These tree-like structures (as shown in Figure \ref{fig:tree_1}) have two main functions. First, they show users the hierarchy of different lessons according to Duolingo's suggestions. Second, they help learners measure and track their progress (according to their learning plans) in a visually appealing way. Each language lesson in the learning tree usually includes several levels of mastery. As students learn and practice a lesson, they progress through these levels of mastery and receive a type of reward called ``crowns.'' Figure \ref{fig:crown} shows an example of crowns won by a user in Duolingo for mastering some lessons. The color of a lesson in the learning tree becomes golden when the user reaches the final level of mastery.
 
However, Duolingo's gamification is not limited to these elements. To maintain the sincere efforts of its learners, Duolingo also uses a particular punitive mechanic known as ``hearts'' to penalize the users who regularly enter too many wrong answers into the system. We should mention here that users usually have a limited number of hearts in their accounts. If they run out of hearts, they will not be able to access more advanced lessons. In such situations, users have to, for example, repeat some old exercises or redo some old lessons to replenish their hearts and regain access to more advanced lessons.

Furthermore, Duolingo keeps its users on their toes with various competitions and gameful challenges. For example, there are currently ten categories of \textit{leagues} in Duolingo: \textit{Bronze}, \textit{Silver}, \textit{Gold}, \textit{Sapphire}, \textit{Ruby}, \textit{Emerald}, \textit{Amethyst}, \textit{Pearl}, \textit{Obsidian}, and \textit{Diamond}. Bronze is the easiest, and Diamond is the most difficult league in Duolingo. In each league, the participating users compete and are ranked on a \textit{leaderboard} based on their weekly earned XP, as shown in Figure \ref{fig:league_leader}. The top-ranked users are then promoted to the higher leagues, and the remaining users enter the ``demotion zone'' and get demoted to lower leagues. Nevertheless, even these demoted users have a chance to redeem their rank and compete again in the next weeks. Duolingo leaderboards and rankings are reset and renewed on a weekly basis.

The second feature that makes Duolingo special is its populated and active \textbf{discussion forums}, where different users can freely express their opinions and ask questions about almost any language-related topic or Duolingo. Although the discussion forums are not directly part of Duolingo's gamification scheme, they provide a rich medium to explore and understand user feedback on gamification. The discussion topics in Duolingo are broadly categorized as follows: \textit{Educators}, \textit{Duolingo}, and \textit{Troubleshooting}, along with the topics specific to each language. This research uses Duolingo's discussion forums to understand ``\textit{what users think about the misuse of gamification.}''

In this appendix, we did our best to summarize and review all of the major features of Duolingo that are necessary for understanding the user quotes reported in this study. However, more detailed information about Duolingo and its features can be found on the application's official website \cite{Duolingo_website_cite}.








\end{document}